# $t'$- and $t''$-dependence of the bulk-limit superconducting condensation energy of the 2D Hubbard model

K. Yamaji[a*], T. Yanagisawa[a], M. Miyazaki[b], R. Kadono[c]

[a] Nanoelectronics Research Institute, AIST, Central 2, 1-1-1 Umezono, Tsukuba 305-8568, Japan

[b] Hakodate National College of Technology, 14-1 Tokura-cho, Hakodate 042-8501, Japan

[c] Institute of Materials Structure Science, KEK, Tsukuba 305-0801, Japan

**Abstract**

The 2D Hubbard model having the 2nd- and 3rd-neighbor transfer energies $t'$ and $t''$ is investigated by use of the variational Monte Carlo method. At the nearly optimal doping with on-site Coulomb energy $U=6$ (energy unit is $t$) the condensation energy $E_{cond}$ for the $d$-wave superconductivity (SC) is computed for lattices of sizes from 10×10 to 28×28 with the aim to get its bulk-limit value. $t''$ is fixed at $-t'/2$. Outside and in the neighborhood of the SDW region of $-0.16 \leq t' \leq -0.08$ the SC $E_{cond}$ dominates over the SDW $E_{cond}$. At $t'=-0.05$ and $-0.10$ we obtained a definitely finite bulk-limit SC $E_{cond}$ of the order of the experimental value for YBCO. At $t'=0$ $E_{cond}$ nearly vanishes. For $t' \lesssim -0.18$, the SC $E_{cond}$ strongly oscillates as a function of the lattice size, when periodic boundary conditions (b.c.'s) are imposed to both axes. In the case of periodic and anti periodic b.c.'s, a finite bulk-limit value is obtained at $t'=-0.22$. $E_{cond}$ tends to vanish with further decrease of $t'$. With our results the SC of LSCO is understandable with $t' \sim -0.10$.

The $t'$ values of Hg1201, Tl2201 and Na-CCOC seem close to –0.20 so that they locate in the boundary zone of SC indicated in the present work. Slightly larger $U$ improves the situation by increasing $E_{cond}$.

*Corresponding author: Nanoelectronics Research Institute, AIST Central 2, 1-1-1 Umezono, Tsukuba 305-8568, Japan; Tel.: +81-298-61-5368; E-mail: yamaji-kuni@aist.go.jp

## 1. Introduction

The mechanism of high $T_c$ superconductivity (SC) in cuprates is still controversial. The first issue for theories based on the electronic interaction is the coupling strength [1-5]. Is it strong, intermediate or weak? Employing moderate on-site Coulomb energy $U$ of about 6 (throughout this paper the energy unit is the nearest neighbor transfer energy $t$), the present authors have been studying if the two-dimensional Hubbard (2DH) model works as the minimal model or not in reproducing the basic properties of high-$T_c$ SC [6-8]. We have computed the SC condensation energy $E_{cond}$ by means of the variational Monte Carlo (VMC) method. At the optimal doping we obtained appropriate values of SC $E_{cond}$ close to the experimental value 0.26 meV per active Cu site, or 0.0007 in unit of transfer energy $t \cong 0.35$eV, of YBCO [6]. Although computations were carried out for large square lattice up to 22×22 [7, 8], there remained an uncertainty about the bulk-limit nature of the SC $E_{cond}$. In the present work we pursue larger lattices up to 28×28 lattices and obtain firm bulk-limit values. Another motivation is to clarify the $t'$ (and $t''$) -dependence of $E_{cond}$ in the context of material dependence of $T_c$, where $t'$ and $t''$ are the second and third neighbor transfer energies, respectively. We have tried to determine the dependence in an appropriately wide range of $t'$ and $t''$ with a fixed relation $t''=-t'/2$ following Ref. [9].

In the present work we restrict ourselves to the optimal doping, or about 0.16/site hole doping, since then the SC $E_{cond}$ takes a relatively large value but should badly suffer from complicating effects neither of the pseudo-gap in the underdoping region nor of the phase separation in the overdoping region. We use 0.16 as the hole doping rate for the continuity to our preceding works, considering the difference from 0.15 not so important due to the broad Bell shape law.

In the next section our model and computation method are briefly given. Results are presented in Section 3 and discussed mainly in the context of high $T_c$ in Section 4.

## 2. Model and Method

Our 2DH model and calculation method are the same as in [8]. The model is defined by

$$H = -t \sum_{<jl>,\sigma} \left(c^\dagger_{j\sigma} c_{l\sigma} + \text{H.c.}\right) - t' \sum_{<<jl>>,\sigma} \left(c^\dagger_{j\sigma} c_{l\sigma} + \text{H.c.}\right)$$
$$-t'' \sum_{<<<jl>>>,\sigma} \left(c^\dagger_{j\sigma} c_{l\sigma} + \text{H.c.}\right) + U \sum_j c^\dagger_{j\uparrow} c_{j\uparrow} c^\dagger_{j\downarrow} c_{j\downarrow}, \tag{1}$$

where $c^\dagger_{j\sigma}$ ($c_{j\sigma}$) is the creation (annihilation) operator of an electron with spin σ at the $j$th site; electronic sites form a rectangular lattice; <jl> denotes the nearest-neighbor pairs; H.c. stands for Hermite conjugate; <<jl>> and <<<jl>>> means the second- and third-neighbor pairs, respectively; *t, t', t"* and $U$ are defined in Introduction. Our trial wave function for the SC state is a modified Gutzwiller-projected BCS-type wave function, which we call Jastrow-type, defined by

$$\Psi_s = \prod_{<jl>} h^{n_j n_l} P_{N_e} \prod_i (1-(1-g)n_{i\uparrow} n_{i\downarrow}) \prod_k (u_k + v_k c^\dagger_{k\uparrow} c^\dagger_{-k\downarrow})|0\rangle, \tag{2}$$

where the third factor operates to vacuum to generate the BCS-type wave function and $c^\dagger_{k\sigma}$ is the Fourier-transformed form of $c^\dagger_{i\sigma}$; the second factor is the Gutzwiller projection operator with $g$ being a variational parameter and $i$ labels a site in the real space; $n_{i\sigma}=c_{i\sigma}^\dagger c_{i\sigma}$; $P_{N_e}$ is a projection operator which extracts the components with a fixed total electron number $N_e$; $h$ is another variational parameter which optimizes the electronic correlation between the nearest neighbor sites. This factor is considerably more effective than a similar one defined for doublon-holon pairs [10a, 10b]. Coefficients $u_k$ and $v_k$ appear in our calculation only in the ratio given by

$u_k/v_k = \Delta_k / \left(\xi_k + \sqrt{\xi_k^2 + \Delta_k^2}\right)$, where $\Delta_k = \Delta(\cos k_x - \cos k_y)$ is a $k$-dependent $d$-wave gap function with gap parameter $\Delta$ and $\xi_k = \varepsilon_k - \mu$ is the band energy $\varepsilon_k = -2t(\cos k_x + \cos k_y) - 4t'\cos k_x \cos k_y - 2t''(\cos 2k_x + \cos 2k_y)$ subtracted by $\mu$, a chemical-potential-like variational variable. The SC ground state energy $E_g = \langle H \rangle \equiv \langle \Psi_s | H | \Psi_s \rangle / \langle \Psi_s | \Psi_s \rangle$ is evaluated by using the Monte Carlo technique. We have minimized $E_g$ by optimizing variational parameters $g$, $h$, $\Delta$ and $\mu$, by use of the correlated measurements method and the Newton method. Condensation energy $E_{\text{cond}}$ for the SC state is obtained as the decrease per site of this $E_g$ from the normal state energy similarly calculated starting from the normal state trial wave function [6]. $E_{\text{cond}}$ for the commensurate SDW state is obtained similarly. The trial SDW wave function given in [6] is only simply modified so that it includes the $t''$ term and the $h$ factor.

## 3. Bulk-limit SC $E_{\text{cond}}$ as a function of $t'$

As descrived in a previous work [8], the SDW $E_{\text{cond}}$ for commensurate wave vector $(\pi, \pi)$ depends on the lattice size $L \times L$ very strongly, where $L$ is the edge length of the square. However, when $L \geq 16$, the size effect subsides. In Fig. 1, the SDW $E_{\text{cond}}$ for $L=20$ is plotted, which can be regarded as the bulk-limit curve in a good approximation. This SDW is presumably due to a kind of nesting in a broad sense of the $k$-regions around the van Hove singularities. This is different from the nesting between straight parts of the Fermi surface at $t'$ close to zero. Due to the latter the SDW $E_{\text{cond}}$ tends to weakly increase with increasing $t'$ around $t'$=0.

As is seen, the SDW $E_{\text{cond}}$ is overwhelmingly large in the region of $-0.16 \leq t' \leq -0.08$. Outside this region it falls. When it subsides, the SC $E_{\text{cond}}$ for L=18 was known to become larger than the SDW $E_{\text{cond}}$ and the SC $E_{\text{cond}}$ may win in the bulk limit as well. This possibility is investigated in the rest of this paper by investigating the SC $E_{\text{cond}}$ for large $L$'s up to 28.

In the region of small $|t'|$ in the r. h. s. of the SDW region in Fig. 1, the results of SC $E_{\text{cond}}$ for large $L$'s up to 28 allow us to obtain a finite bulk-limit value in a clear-cut way. In Fig. 2 the SC $E_{\text{cond}}$ in the case of $t'=-2t''=-0.05$ is plotted against $1/L^2$ up to $L=28$. The $E_{\text{cond}}$ value and error bars are the average and the standard deviation, respectively, defined with the $E_{\text{cond}}$ values computed in many parallel CPU units. The periodic b. c.'s for both axes are employed. The obtained SC $E_{\text{cond}}$ has a very clear tendency to converge to a finite value, *i. e.*, bulk-limit value, as $1/L^2$ goes to zero. A linear fit is obtained for $L$=12 ~ 28 with the bulk-limit value $E_{\text{cond}}$ =0.0014.

In Fig. 3 the SC $E_{\text{cond}}$ in the case of $t'=-2t''=-0.10$ is plotted similarly. Again we get definitely a bulk limit. A linear fit with $L=14 \sim 28$ leads to the bulk-limit value $E_{\text{cond}}$ =0.0010. In the case of $t'=-2t''=-0.10$, the SDW $E_{\text{cond}}$ is predominant over the SC $E_{\text{cond}}$ so that the pure SC state is not realized but, as our preceding work indicated by a limited size computation, an incommensurate SDW with $(\pi\pm\delta, \pi)$ or $(\pi, \pi\pm\delta)$ is stabilized with which an SC ordering coexists [11]. We expect such a state occurs in the bulk-limit in the SDW-dominant region.

In Fig. 3 the SC $E_{\text{cond}}$ in the case of $t'=t''=0$ is plotted as well. Here it decreases in an unexpected way when $L$ exceeds 20. It indicates the bulk-limit vanishes or is less than 0.0001, although it is difficult to definitively state since error bars are large.

Together with the results for $t'=-0.05$ we can say that the SC $E_{\text{cond}}$ has a tendency to increase with increasing $|t'|$ at least up to 0.05.

A typical result in the region of $t'\leq-0.15$ is shown in Fig. 4 with $t'=-2t''=-0.22$. As is seen, the SC $E_{\text{cond}}$ is oscillatory as a function of $L$, having peaks at $L=12$ and 24. From the plot we cannot get a straight-forward extrapolation to the limit of $L=\infty$. The oscillatory peaks of $E_{\text{cond}}$ are seen to shift systematically as a function of $t'$ $(=-2t'')$ as seen in Fig. 5. The peak at $L=24$ in Fig. 4 moves to $L\sim28$ in the plot for $t'=-0.20$ and goes beyond the range of up to $L=28$ in the plot for $t'=-0.18$. On the other hand another series of peaks are seen at $L=24$, 20 and 18 in the plots for $t'=-0.15$, $-0.18$ and $-0.20$, respectively. The size effect is considered to arise from biased distributions of the one-particle energy $\varepsilon_k$. As already explained [8], the peaks and bottoms in the plots appear since $\varepsilon_k$ has flat parts of the Fermi surface nearly parallel to the $k_x$ or $k_y$ axis, when $|t'|$ is large. Due to this feature $\varepsilon_k$'s allotted at reciprocal lattice points are condensed in energy scale into some groups, in each of which the $k$ points are arranged on a $k_x=$const (or $k_y=$const) line nearly parallel to a flat part of the Fermi surface and have $\varepsilon_k$'s close to each other. When the Fermi energy $\varepsilon_F$ lies in the midst of such one group, $E_{\text{cond}}$ is enhanced. With $\varepsilon_F$ happens to lie in between two groups, $E_{\text{cond}}$ becomes small. For this reason the oscillatory behavior occurs at large values of $L$ albeit oscillation amplitude should fade away at very large $L$ values. In the cases of $-0.22\leq t'\leq-0.18$ the average value of $E_{\text{cond}}$ is guessed by sight to be 0.0004 or less.

When the periodic b. c.'s for an axis and antiperiodic one for the other, the number of $\varepsilon_k$'s with different values doubles so that the oscillation of $E_{\text{cond}}$ should weaken. In Fig. 4 diamond symbols meaning $E_{\text{cond}}$ for these b. c.'s satisfy the expectation and give a finite bulk-limit value 0.00019, although this value is much smaller than the typical value $E_{\text{cond}} \cong 0.0014$ at $t'=-0.05$ on the other side of the $t'$ region.

In Ref. [8] we examined the effect of b. c.'s to oscillatory $E_{\text{cond}}$ with $L$ up to 20 for $t'=-0.31$ and $t''=0.21$ and found that the average is less oscillatory. The computation with the periodic b. c.'s is extended for $L$ up to 28 with the results in Fig. 6. On the basis of a reasoning similar as given above, we expect to have another appreciable peak at $L \approx 28$ but it proves very small. Thus we judge the bulk-limit value for this set of $t'$ and $t''$ is less than 0.0001 or more plausibly vanishes.

The error $\delta(E_{\text{cond}})$ brought about to $E_{\text{cond}}$ by the deviation $|\rho-0.84|$ of the employed electron density $\rho=N_e/L^2$ from the theoretical 0.84 is given by $|\rho-0.84| \times |\text{d(intrinsic } E_{\text{cond}})/\text{d}\rho|$. $|\rho-0.84|$ is smaller than $1/L^2$, decreasing with increasing $L$. It becomes further smaller; actually the maximum deviation is 0.00116 when $18 \leq L \leq 28$. To estimate d(intrinsic $E_{\text{cond}})/\text{d}\rho$ we have old data for $\rho=0.80, 0.82$ and 0.84 with $t'=-0.10$ and $t''=0$ ; $U=8$ and $h$ is fixed at 1. Neglecting size effect fluctuations, we get d(intrinsic $E_{\text{cond}})/\text{d}\rho \sim 0.0203$ as an overestimation. The effects of different $U$ and fixed $h$ almost cancel each other. Thus in the neighborhood of $t'=-0.10$ we get $\delta(E_{\text{cond}}) < 0.000023$. With $t'=-2t''=-0.20$ our recent tentative data suggest that d(intrinsic $E_{\text{cond}})/\text{d}\rho$ takes a smaller absolute value with a negative sign so that $\delta(E_{\text{cond}}) < 0.000023$ is valid around here as well. Since it is smaller than $E_{\text{cond}}$ by almost two orders of magnitude, $\delta(E_{\text{cond}})$ can be safely neglected.

The results for the bulk-limit values as a function of $t'$ are plotted in Fig. 1.

The SC $E_{\text{cond}}$ is seen to increase as $t'$ decreases from zero to the values in the SDW region in Fig. 1. After $t'$ goes through it, the SC $E_{\text{cond}}$ diminishes and deceases to ~0 with further decreasing $t'$. This is natural since the SC $E_{cond}$ is considered to be given rise due to the enhanced spin susceptibility [5] due to a certain sort of Fermi surface nesting. Similar enhancement and decreasing of $T_c$ in the vicinity of the SDW region were reported by use of the FLEX theory for the anisotropic 2DH model in which the linking second-neighbor sites by $t'$ is restricted only in one diagonal direction [12].

For the strong coupling model, *i. e.*, t-J model with $t'$ and $t''=-t'/2$, the SC correlation was reported to increase with decreasing $t'$ from zero to about –0.30 and then to decrease by use of a VMC calculation [13]. The difference of the transient $t'$ value may from the smaller lattice size ($L=12$) as well as the difference of the model.

## 4. Is high $T_c$ in cuprates understandable?

The experimental SC $E_{cond}$ for LSCO is estimated at 0.029meV/(Cu site), or 0.00008 in unit of $t$ [14]. The band parameter values of LSCO was estimated at $t'=-0.12$ and $t''=0.08$ [15]. This set corresponds roughly to the $t'\cong-0.10$ point in Fig. 1, for which $E_{cond}\cong0.0010$. The latter value is much larger than the above-mentioned experimental value. However, the SDW $E_{cond}$ is much larger here. Our preceding work [11] showed that in this situation, an incommensurate, more precisely, stripe-type SDW state coexisting with SC has a lower total energy and that the SC part of the whole $E_{cond}$ is much reduced. Therefore, such a coexistence allows from us to qualitatively understand the SC $E_{cond}$ in LSCO.

Concerning the one-$CuO_2$-layer cuprates, Tl2201 ($T_c=93K$) and Hg1201 ($T_c=98K$) band calculations by Singh et al. [16] give very much deformed Fermi surfaces which can be fitted with $t'\sim-0.40$, which is a hopeless situation for the present scheme to give SC. However, there are some experimental results reporting the forms of the Fermi surface. In the case of Tl 2201, an AMRO work [17] gives such information which allows to get $t'\cong-0.20$ and $t''\cong 0.165$. There is also an ARPES report [18], which provides similar values. In the case of Hg1201, there is an ARPES report [19], from which we obtain by fitting $t'\cong-0.20$ and $t''\cong 0.175$. There are also reports of the Fermi surface for Bi2201 [20] and (Ca,Na)COCl [21]. The value of $|t'|$ or $t''$ is slightly smaller in both materials.

These sets of $t'$ and $t''$ are located in the boundary zone of the SC region in Fig. 1. Here the calculated SC $E_{cond}$ is smaller by a factor around 4 than the experimental one ~0.0007 for YBCO even when the detrimental effect of the pseudogap is neglected. From our viewpoint of the 2DH model, this discrepancy could be filled by slight increase of $U$. Actually when $U$ is increased to 7, a preliminary computation shows a quick increase of $E_{cond}$ in the transient region of $t'$. Also slight increase of hole doping rate may lead to a larger $E_{cond}$, as suggested by a negative value of d(intrinsic $E_{cond}$)/d$\rho$ around $t'=-2t''=-0.20$ in Section 3. This may mean that the optimal doping is slightly dependent on the band parameter values.

The main results are summarized in Abstract and Fig. 1.

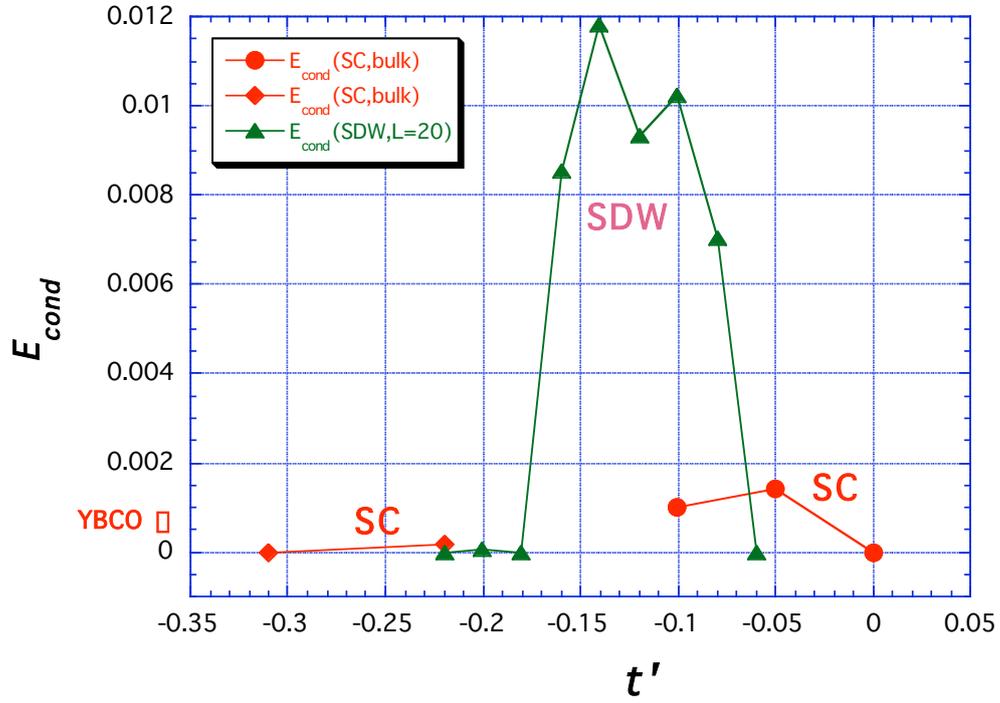

Fig. 1. Triangles give the SDW $E_{cond}$ with $L$=20, indicating the SDW-dominant region. Obtained bulk-limit SC $E_{cond}$ is plotted against $t'$ for $t'$=0, −0.05, −0.10 and −0.22 with circles and diamonds; $t''$ is fixed at $-t'/2$. The diamond for $t'$=−0.31 is exceptionally with $t''$=0.21. Electron density $\rho$ is nearly optimal with $\rho \cong 0.84$ and $U$ is fixed at 6. YBCO in the figure indicates the experimental SC $E_{cond}$ for $YBa_2Cu_3O_7$.

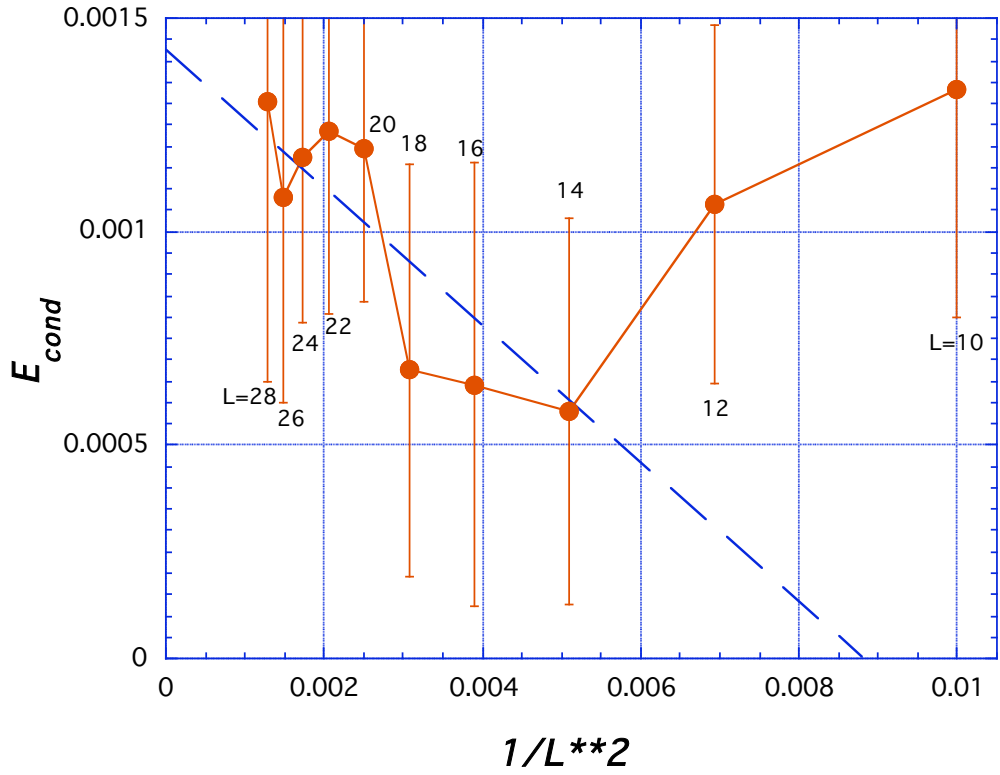

Fig. 2. SC $E_{cond}$ plotted against $1/L^2$ for $t'=-0.05$ ($t''=-t'/2$) with $U=6$ and $\rho \cong .84$. $L$ is the lattice edge length. The dashed line is a linear fit with $L=12\sim28$, giving the bulk-limit value 0.0014. The boundary conditions are periodic for both axes.

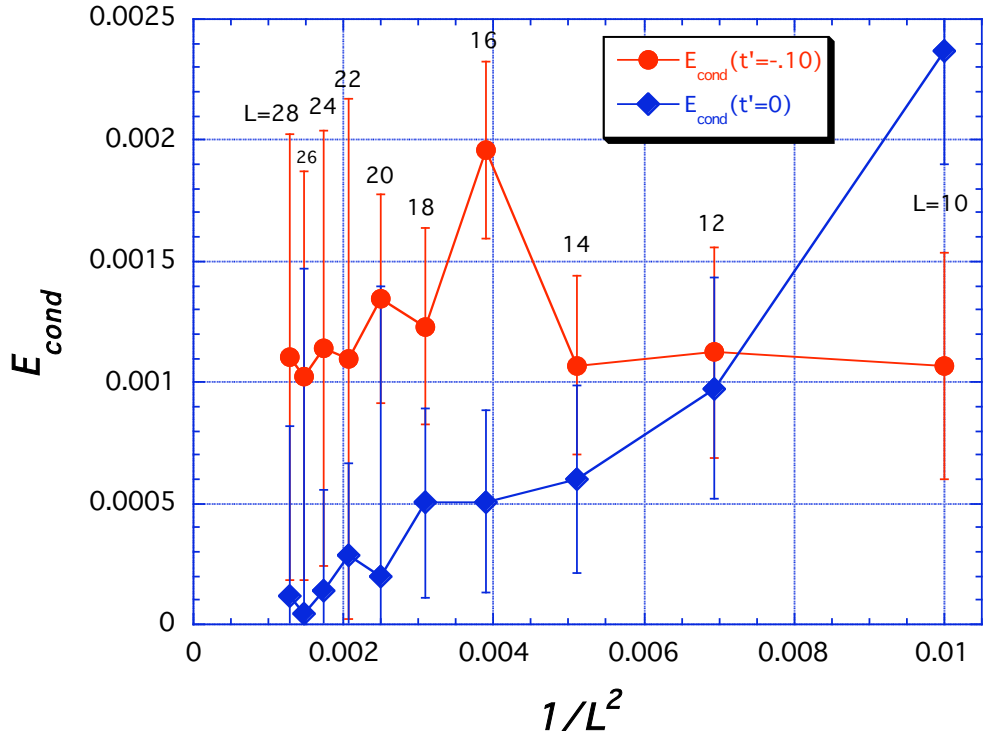

Fig. 3. SC $E_{cond}$ is plotted against $1/L^2$ for $t'=-0.10$ with circles. A linear fit with $L=14\sim 28$ gives the bulk-limit value 0.0010 (not drawn). SC $E_{cond}$ is plotted against $1/L^2$ also for $t'=0.0$ with diamonds. Other settings are the same as in Fig. 2.

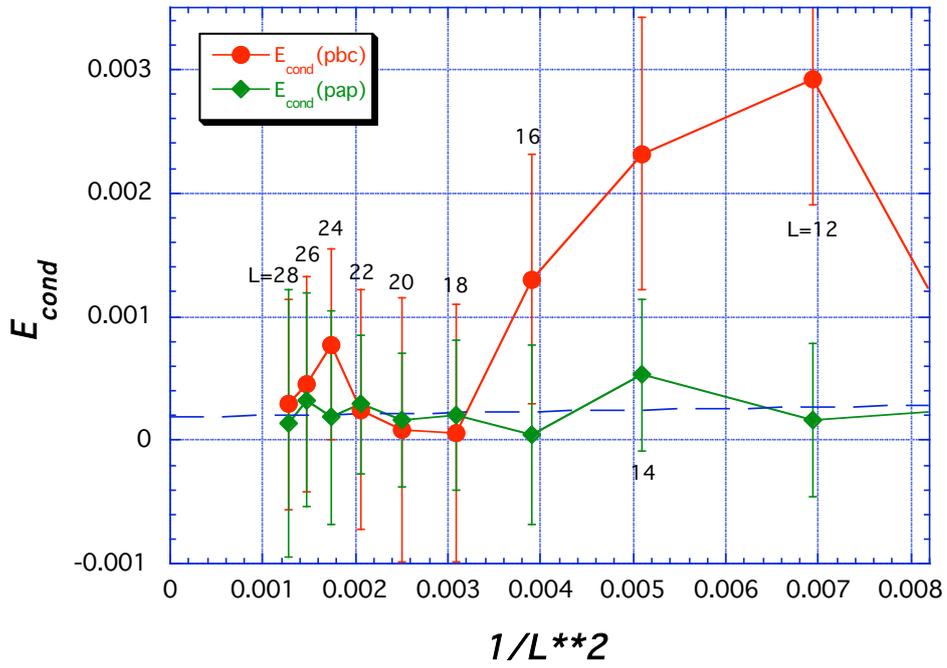

Fig. 4. SC $E_{cond}$ plotted against $1/L^2$ for $t'=-0.22$ with circles in the case of the periodic b. c.'s. Diamonds denote the SC $E_{cond}$ calculated in the case of the periodic and antiperiodic b. c.'s. The linear fit gives the bulk-limit 0.00019. Other settings are the same as in Fig. 2.

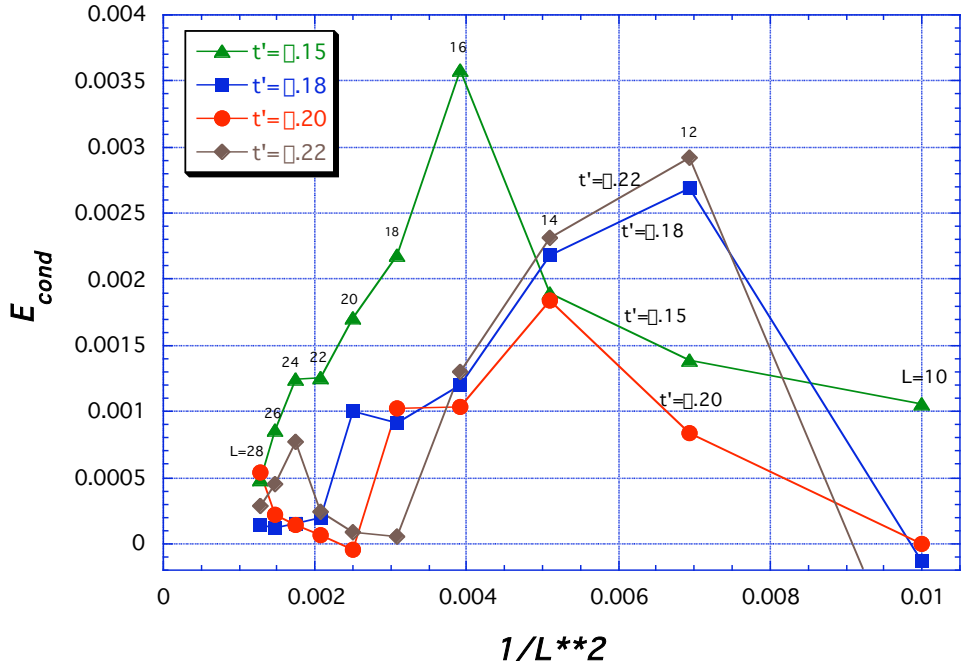

Fig. 5. SC $E_{\text{cond}}$ plotted against $1/L^2$ for $t'$=−0.15 (triangle), −0.18 (square), −0.20 (circle) and −0.22 (diamond). Other settings are the same as in Fig. 2.

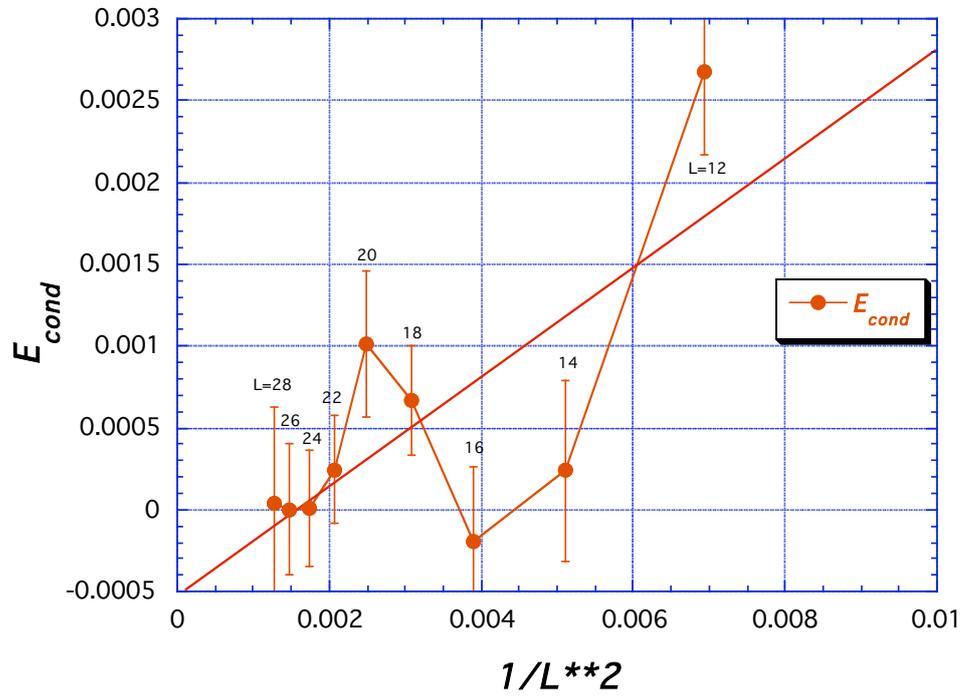

Fig. 6. SC $E_{\text{cond}}$ plotted against $1/L^2$ for $t'=-0.31$ and $t''=0.21$. Other settings are the same as in Fig. 2.